\documentclass[twocolumn,showpacs,preprintnumbers,amsmath,amssymb]{revtex4}

\usepackage{graphicx}
\usepackage{dcolumn}
\usepackage{bm}

\begin{document}


\title{Modeling Study of the Dip/Hump Feature \\
in Bi$_2$Sr$_2$CaCu$_2$O$_{8+\delta}$ Tunneling Spectroscopy}

\author{P. Romano$^{1}$}
\author{L. Ozyuzer$^{2}$}
 \email{ozyuzer@iyte.edu.tr}
\author{Z. Yusof$^{3}$}
\author{C. Kurter$^{4}$}
\author{J. F. Zasadzinski$^{4}$}
\affiliation{%
$^{1}$CNR-INFM SUPERMAT Salerno and DSBA, Via Port'Arsa 11, Univ. del 
Sannio, I-82100 Benevento, Italy
}%
\affiliation{
$^{2}$Department of Physics, Izmir Institute of Technology, TR-35430 Izmir, Turkey}%
\affiliation{$^{3}$High Energy Physics Division, Argonne National Laboratory, Argonne, IL 60439}
\affiliation{$^{4}$Physics Division, Illinois Institute of Technology, Chicago, IL 60616
}

\date{\today}

\begin{abstract}
The tunneling spectra of high temperature superconductors on 
Bi$_2$Sr$_2$CaCu$_2$O$_{8+\delta}$
(Bi-2212) reproducibly show a high bias structure in the form of a dip-hump at 
voltages higher than the gap voltage. Of central concern is whether this feature 
originates from the normal state background or is intrinsic to the 
superconducting mechanism. 
We address this issue by generating a set of model conductance curves - a 
''normal state'' 
conductance that takes into account effects such as the band structure and a possible 
pseudogap, and a pure superconducting state conductance. When combined, the 
result shows that the dip-hump feature present in the experimental conductance curves 
cannot be naively attributed to a normal state effect. In particular, 
strong dip features found in 
superconductor-insulator-superconductor data on optimally-doped Bi-2212, including 
negative dI/dV, cannot be a consequence of an extrinsic pseudogap. However, 
such features can easily arise from states-conserving deviations in the superconducting 
density of states, e.g., from strong-coupling effects.    

\end{abstract}

\maketitle

Tunneling spectroscopy of conventional superconductors produces a direct, 
high-resolution measurement of the quasiparticle density of states (DOS). 
The fine structures seen in such measurements have been linked directly to 
the superconducting mechanism in the conventional superconductors \cite{wolf}.  
The same technique has also been applied to high-T$_{c}$ superconductors (HTS).  
Among HTS cuprates, Bi$_2$Sr$_2$CaCu$_2$O$_{8+\delta}$ (Bi-2212) 
is one of the most extensively studied by tunneling and other surface 
sensitive experiments. This is due to the ability to easily cleave the single crystals along the a-b 
plane as well as possibility of obtaining large single crystals with various doping concentrations. 
Unfortunately, the sheer diversity of tunneling spectra from tunneling measurements have led to a
 variety of interpretations of the spectral features. One example is the debate 
on the nature of a dip-hump structure observed in the tunneling spectra. The dip is a 
strong deviation from tunneling conductance beyond the quasiparticle peak, which is followed 
by a hump first observed in 1989 \cite{huang89}. There are two school of thoughts on the origin of this structure. 
The first is that, similar to what has been learned from conventional superconductors, this feature 
is intrinsic to the superconducting mechanism and is analogous to the phonon fine structure \cite{huang90}. 
Thus, a thorough analysis of tunneling spectra including this feature should produce the necessary 
parameters in unveiling the nature of superconductivity in HTS. The other argument points to idea 
that the dip-hump feature is simply a normal state or background effect independent of the 
superconducting mechanism. The explanations range from the hump being the remnant of the 
van Hove singularity (VHS) to the dip as being due to the presence of the pseudogap 
\cite{bouvier,yurgens99,suzuki}.  
This means that the hump is the pseudogap peak while the dip is simply 
the ''valley'' between the 
pseudogap peak and the superconducting peak.  Under this assumption, the dip-hump feature is 
a distraction from the intrinsic superconducting properties and can be normalized away from the 
tunneling conductance. We will address this last point by generating a model conductance curve 
that includes the normal state features and show that these normal state effects are insufficient in 
reproducing the salient characteristics of the dip-hump structure, especially the very strong dip 
feature found in break junctions.  In this study, the presented tunneling conductance of break 
junctions in Bi-2212 is an extreme case that show even negative tunneling conductance in the dips voltage. 
We show using a simple model of strong coupling effects, that such strong dips are 
easily obtained. Moreover, using neutron scattering 
data, a model is also suggested by Eschrig and Norman \cite{eschrig} which can produce negative tunneling 
conductance in dips of superconductor-insulator-superconductor (SIS) junctions. 
This strongly suggests that the dip structure in the tunneling conductance spectra 
has a superconducting origin, and is likely tied to the mechanism of pairing.  

\begin{figure}
\vspace{-0.7in}
\centering
\includegraphics[bb=40 55 540 720, width=3in]{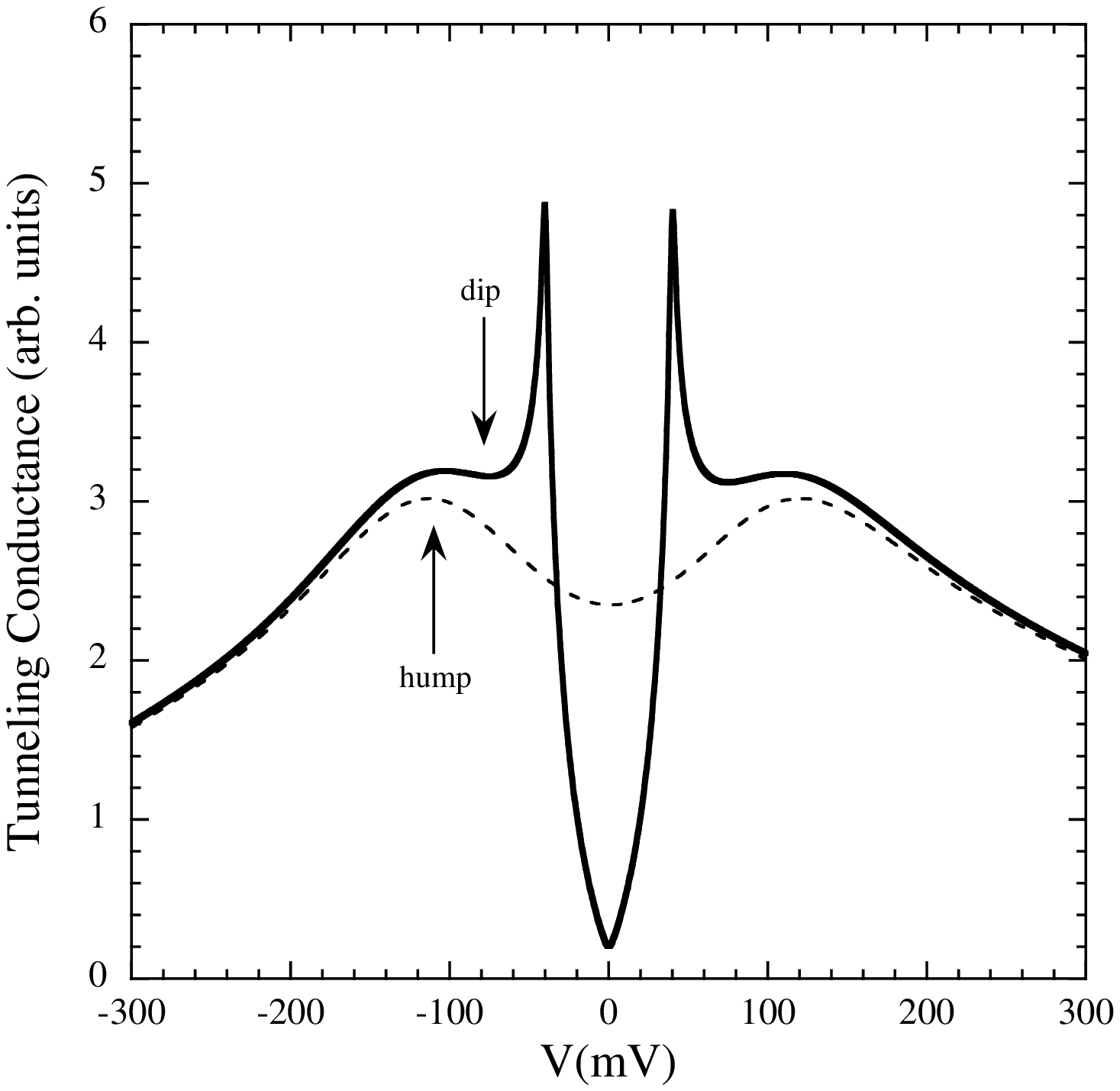}
\vspace{-0.9in}
\caption{\label{fig:epsart} Dashed line: normal state DOS generated using  $\Delta_{p}$ = 120 meV and $\Gamma_{p}$ =90 meV.
	Solid line: the model tunneling DOS using the normal state DOS and superconducting DOS (Won and Maki model). }
\end{figure}

We review and consider tunnel junctions that have been fabricated on Bi-2212, using 
Scanning Tunneling Microscopy/Spectroscopy, point contact, break junctions, and fully 
intercalated intrinsic c-axis junctions \cite{yurgens99,dewilde,renner,ozyuzer00}.  
Since SIS data enhances spectral features, 
we have focused specifically on break junction experiments in this study.  
We do not include intrinsic Josephson junctions (IJJ)  because they are severely affected by joule heating \cite{zavaritsky}. 
For such junctions, the intercalation of Bi-2212 with 
HgBr$_{2}$ increases the barrier thickness between the CuO$_{2}$ layers and reduces the heating caused 
by overinjection \cite{kurter}.   Only then are the dip and hump features recovered in the intrinsic density of states. 

The main features commonly observed in the tunneling 
conductance spectra of break junctions are (i) energy gap at voltages $\pm$ 2$\Delta$/e with $\Delta$ ranging from as low as 10 meV 
for highly overdoped to as high as 60 meV for highly underdoped samples and (ii) high bias structures 
in the form of dip and hump at voltages higher than $\pm$ 2$\Delta$/e. The strong dependence of the magnitude of 
energy gap on doping concentration is well known \cite{miyakawa99}.  
For optimally doped samples of Bi-2212, the 
value of the energy gap is now understood to be about 37 meV. The overall shape of tunneling spectra 
is also very reproducible, and for this reason Bi-2212 is a good candidate to understand the mechanism 
of superconductivity in the cuprates.

The reproducible presence of the dip-hump just above the gap feature has been 
observed not only in Bi-2212, but also in Tl$_{2}$Ba$_{2}$CuO$_{6+ \delta}$ \cite{ozyuzer99} 
and (Cu,C)Ba$_{2}$Ca$_{3}$Cu$_{4}$O$_{12+ \delta}$ cuprates \cite{miyakawa03}. In SIS junctions, the 
dip appears at a voltage (2$\Delta$+$\Omega$)/e, where the mode energy, $\Omega$,  is 
proportional to T$_{c}$ \cite{zasadzinski01}.  
Arguments have been made that this dip-hump structure is purely non-superconducting in 
nature \cite{markiewicz}.  It is the result of the presence of two energy gaps in the total DOS, one 
from the superconducting gap in the superconducting DOS, and the other from the pseudogap, 
which is present even above T$_{c}$, and thus, is considered as part of the normal state DOS effect.
 It has been argued that the product of these two DOS should produce the identical dip-hump 
structure seen in the tunneling conductance. If this is true, then in principle, one can normalize the 
conductance data to obtain the pure superconducting DOS, a practice that has been successfully 
employed for the conventional superconductors. The focus of this paper is to investigate such claim. 
We do this by generating a model conductance data that is a product of 
a ''pure'' superconducting 
DOS having the superconducting energy gap and quasiparticle peaks, and a purely normal state 
DOS incorporating the band structure and an extrinsic pseudogap. 

\begin{figure}
\vspace{-0.7in}
\centering
\includegraphics[bb=40 55 540 720, width=3in]{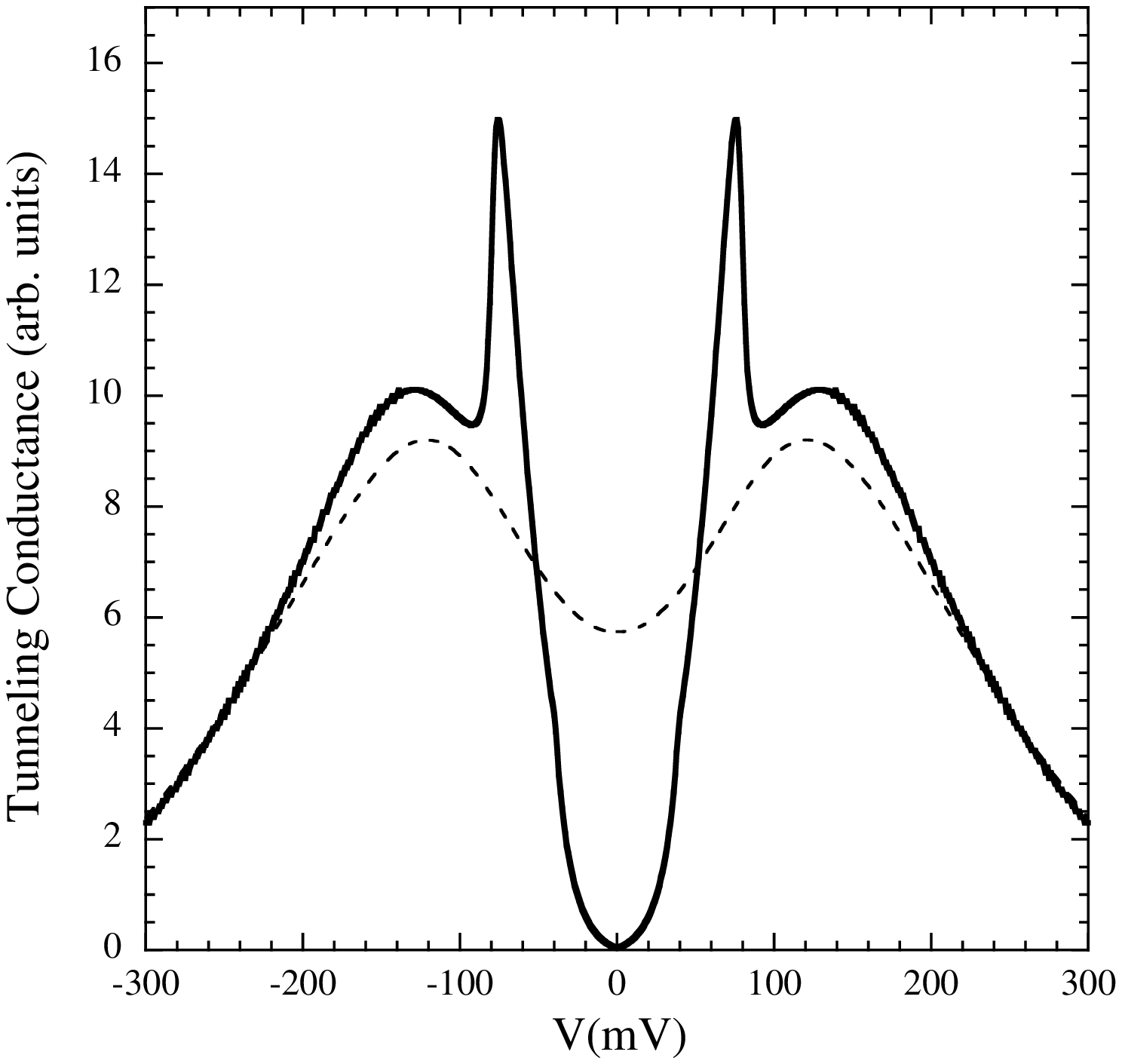}
\vspace{-0.9in}
\caption{\label{fig:epsart} SIS tunneling DOS generated by convoluting the model DOS from Fig. 1 with itself. }
\end{figure}

In principle, at zero temperature, the tunneling conductance from a superconductor-insulator-normal metal (SIN) 
junction gives a direct measure of the DOS \cite{wolf}.  In conventional superconductors, a normalization 
procedure of the tunneling conductance removes the normal state effects. 
What is left behind is essentially the superconducting DOS that has been shown to match the 
BCS DOS. Therefore, to generate our model conductance for comparison with the experimental 
SIN tunneling conductance, we need to generate two separate components, the normal state DOS 
and the superconducting DOS. The normal state DOS is obtained from the momentum-averaged single 
particle Green's function, i.e.  \( N(E)= \frac{1}{\pi} \sum_{k} ImG(k,E) \).  Our model incorporates 
several important 
parameters: (i) a realistic band structure for the CuO$_{2}$ plane that is derived from ARPES measurement 
on Bi-2212 \cite{norman}; and (ii) $d$-wave pseudogap order parameter 
$\Delta_{p}$($k$)=$\Delta_{p}$ [cos($k_{x}$a) - cos($k_{y}$a)]/2 
to include 
the presence of any normal-state gap (pseudogap). The pseudogap is modeled by introducing a very large 
scattering rate $\Gamma$$_{p} \sim \Delta$$_{p}$. The model is described in detail in Ref. \cite{yusof}. 
In this work, we assume that the 
tunneling matrix is constant.  The superconducting DOS is obtained from the model used by Won and 
Maki \cite{won},

\begin{equation}
\sigma (E) = Re 
\left\{ 
\frac{E-i \Gamma_{s}}{\sqrt{(E-i \Gamma_{s} )^{2}- \Delta_{s} (k)^2}}
\right\}
\label{eq:one}
\end{equation}
\noindent
with $\Delta$($k$)=$\Delta$cos(2$\phi$), i.e. having the same symmetry as 
the $d$-wave gap 
with $\phi$ 
measured from the Cu-O bonding direction in the a-b plane. 
Here, $\Gamma_{s}$ is the lifetime 
broadening term for superconducting quasiparticles and is 
assumed small, $\Gamma_{s}$$\ll$$\Delta_{s}$. 
To account for the anisotropic band structure in the cuprate layer, a small directionality factor 
has been included to the superconducting DOS, \cite{ozyuzer00} which provides a better fit of the sub-gap 
region of high quality STM data \cite{zasadzinski03}.

A ''normal state'' DOS has been generated with $\Gamma$$_{p}$ = 90 meV and 
a $d$-wave pseudogap magnitude 
$\Delta$$_{p}$= 120 meV as shown in dashed line in Fig.\ 1. These 
parameters give a qualitative agreement to the estimated 
background conductance from a SIN data at high positive bias.  It reproduces the often-seen weakly decreasing 
background in most SIN spectra. The pseudogap also broadens the van Hove singularity found in the negative 
bias side of the DOS, making it undetectable.  Next, a superconducting $d$-wave DOS has then been generated 
using Eq.\ 1 with a $\Delta_{s}$ = 30 meV, $\Gamma_{s}$ = 1.5 meV. These 
parameters were chosen to reproduce the typical 
SIN data in the gap region, which includes the shape of the tunneling gap and the quasiparticle peaks 
often seen in the experimental data \cite{ozyuzer00,ozyuzer99,zasadzinski01,zasadzinski03}.

The product of normal state DOS and the superconducting DOS produces the tunneling DOS, 
which is shown as the solid curve in Fig.\ 1. This model tunneling 
conductance appears to have the 
generic features of a typical SIN data, including the dip-hump structure. We proceed a step further 
by using the model conductance to generate a SIS conductance. 
This is shown in Fig.\ 2 along with the 
convoluted background conductance.  The most important observation of the model SIS curve is that, even 
though the dip-hump feature is more pronounced, the minimum of the dip is never below the background value.  
This is a crucial outcome of the model that incorporates the pseudogap as part of the normal state background, 
and is contradictory with temperature dependent tunneling studies \cite{yurgens99,miyakawa99}, which clearly show that the dip 
minimum is far below the normal state conductance.

\begin{figure}
\vspace{-0.7in}
\centering
\includegraphics[bb=40 55 540 720, width=3in]{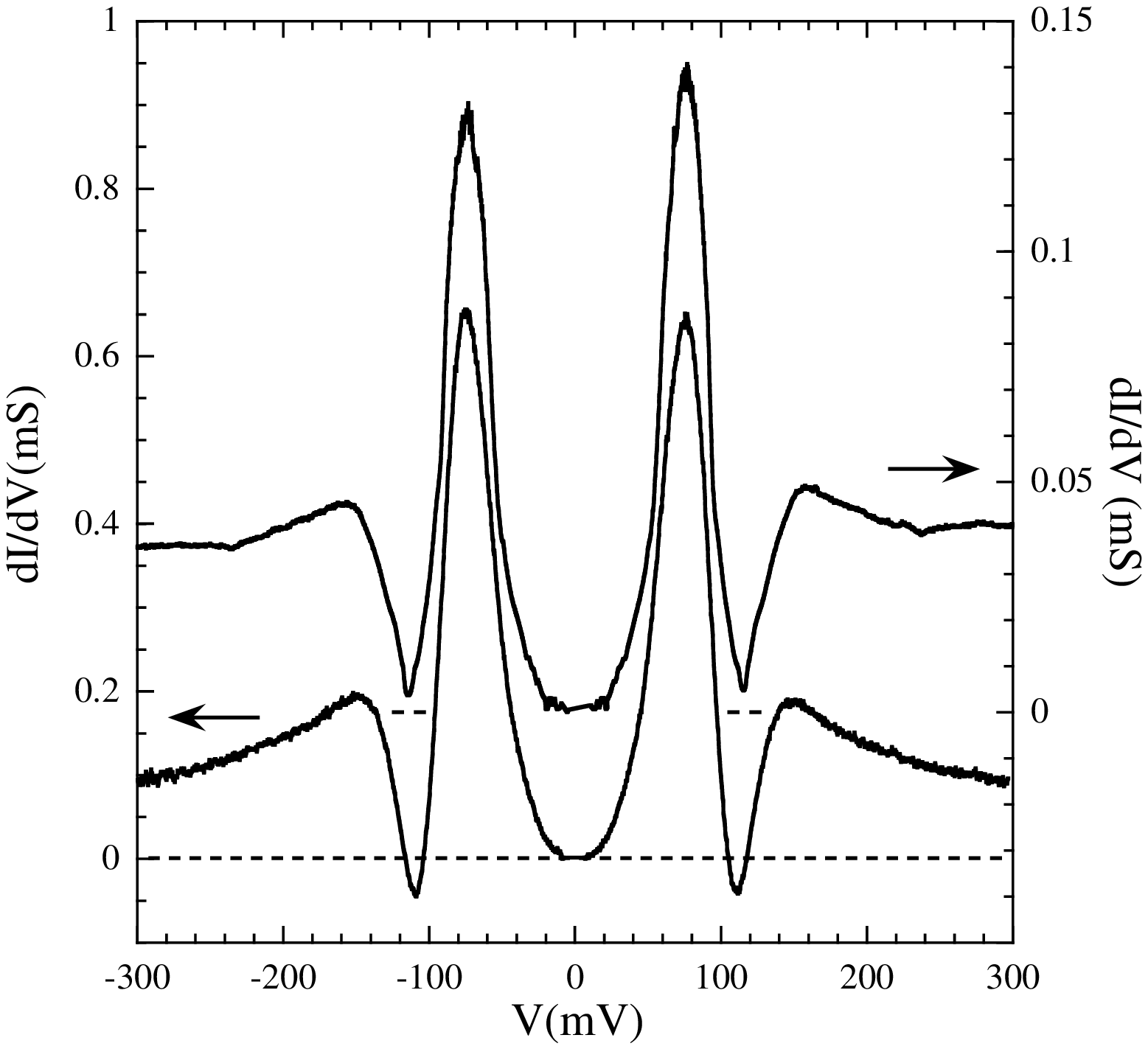}
\vspace{-0.9in}
\caption{\label{fig:epsart} Experimental SIS tunneling data of optimally doped Bi-2212 at 4.2 K.  
Note the severe drop of the dip amplitude, reaching even negative values. }
\end{figure}

We demonstrate the significant discrepancy between the model SIS 
curve of Fig.\ 2 and the two experimental 
SIS data shown in Fig.\ 3 \cite{miyakawa99}.  The two curves are from two 
different optimally-doped Bi-2212 crystals obtained 
at 4.2 K.  Josephson current peaks have been deleted for clarity.  Both experimental data presented here are the 
extreme cases where the dip minima are close to zero.  In fact, the lower SIS curve even shows negative 
conductance values for the dip minima.  Such features cannot be found from the model conductance curve.  
We conclude that attributing the dip as simply the valley between an extrinsic pseudogap peak and the 
superconducting peak is incompatible with experimental observations.  

\begin{figure}
\vspace{-0.6in}
\centering
\includegraphics[bb=120 55 540 720, width=4in]{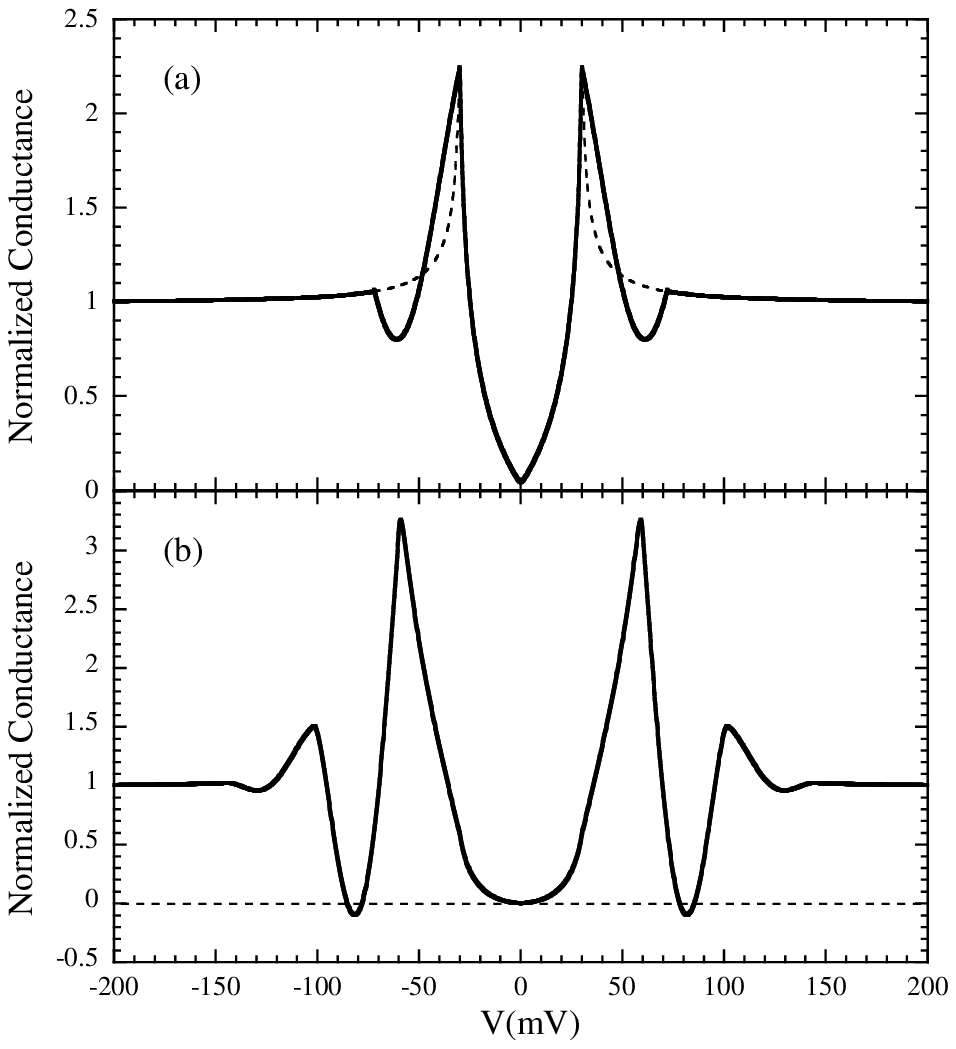}
\vspace{-2.9in}
\caption{\label{fig:epsart} (a) The dip is introduced as a removal of states from the model superconducting 
DOS (solid line).  The dashed line is $d$-wave model without the dip. (b) Self-convolution produces the model SIS 
DOS. It has the identical feature of the dip-hump seen in the experimental data of Fig. 3. }
\end{figure}

On the other hand, if the dip is incorporated into the superconducting DOS in a similar manner as the 
phonon structure seen in the tunneling conductance of conventional superconductor \cite{huang90}, then all aspects 
of the experimental dip-hump feature can be reproduced.  Fig.\ 4(a) shows 
the dip as a removal of states 
from the unperturbed (dashed curve) BCS superconducting DOS.  To conserve states, an excess or 
pile-up of states is placed near the gap edge.  This was obtained by multiplying the $d$-wave DOS by one 
wavelength of a sine function, which guaranteed the number of states would be conserved.  The resulting 
model DOS was then convoluted to again generate the model SIS tunneling conductance.  As shown in 
Fig.\ 4(b), the dip minimum is significantly enhanced and even reaches 
negative values.  The dip clearly extends below the background in contrast 
to the model conductance of Fig.\ 1. This produces a better agreement 
with the experimental observations of Fig.\ 3.

The dip-hump feature observed in tunneling conductance experiment cannot be accurately 
attributed to normal state effects of an extrinsic pseudogap.  The model conductance 
incorporating such normal state effects, while producing a dip-hump structure, cannot 
accurately reproduce the important characteristics of this structure, especially in the SIS 
tunneling spectra.  In particular, the pseudogap model shows that the dip minimum is 
always lies above the normal state conductance, a feature that is contradicted by several experiments.  
More importantly, the model can never produce a negative dynamic conductance, an effect which has 
been observed experimentally.  On the other hand, a dip which drops below the normal state and 
produces a negative dI/dV, are effects that are easily reproduced and understood within a framework 
of a states-conserving deviation from a BCS DOS, as is found with strong-coupling effects.

\begin{acknowledgments}
L.O. acknowledges support from the Turkish Academy of Sciences, in the framework of the 
Young Scientist Award Program (LO/TUBA-GEBIP/2002-1-17).
\end{acknowledgments}


\begin{references}

\bibitem{wolf} E. L. Wolf {\em Principles of Electron Tunneling Spectroscopy }
(Oxford University Press, Oxford, 1985) 

\bibitem{huang89} Q. Huang, J. F. Zasadzinski, K. E. Gray, J. Z. Liu, H. Claus, {\em Phys. Rev. B} {\bf 40},
R9366 (1989)

\bibitem{huang90} Q. Huang, J. F. Zasadzinski, N. Tralshawala, K. E. Gray, 
D.G. Hinks, J. L. Peng, R.L. Greene, {\em Nature (London)} {\bf 347},
369 (1990)

\bibitem{bouvier} J. Bouvier and J. Bok, {\em  Physica C } {\bf 288},
217 (1997)


\bibitem{yurgens99} A. Yurgens, D. Winkler, T. Cleason, S. Hwang and J. Choy, 
{\em Int. J. Mod. Phys. B} {\bf 13}, 3758 (1999)

\bibitem{suzuki} M. Suzuki and T. Watanabe, {\em Phys. Rev. Lett.} {\bf 85},
4787 (2000)

\bibitem{eschrig}  M. Eschrig and M. R. Norman, {\em Phys. Rev. Lett.} {\bf 85},
3261 (2000)

\bibitem{dewilde} Y. DeWilde, N. Miyakawa, P. Guptasarma, M. Iavarone, L. Ozyuzer, 
J.F. Zasadzinski, P. Romano, D. G. Hinks, C. Kendziora, G. W. Crabtree and 
K. E. Gray, {\em Phys. Rev. Lett.} {\bf 80},
153 (1998)


\bibitem{renner} Ch. Renner, B. Revaz, J.Y. Genoud, K. Kadowaki and O. Fischer, {\em Phys. Rev. Lett.} {\bf 80},
149 (1998)

\bibitem{ozyuzer00} L. Ozyuzer, J.F. Zasadzinski, C. Kendziora, K. E. Gray, {\em Phys. Rev. B} {\bf 61},
3629 (2000)

\bibitem{zavaritsky} V. N. Zavaritsky, {\em Phys. Rev. B} {\bf 72},
094503 (2005)


\bibitem{kurter} C. Kurter, L. Ozyuzer, J.F. Zasadzinski et al., {\em unpublished}; L. Ozyuzer, C. Kurter, J. F. Zasadzinski, K. E. Gray, D. G. Hinks, 
N. Miyakawa, {\em IEEE Trans. Appl. Supercond.} {\bf 15}, 181 (2005).   


\bibitem{miyakawa99} N. Miyakawa, J. F. Zasadzinski, L. Ozyuzer, P. Guptasarma , 
D. G. Hinks, C. Kendziora and K. E. Gray, {\em Phys. Rev. Lett.} {\bf 83},
1018 (1998)

\bibitem{ozyuzer99} L. Ozyuzer, Z. Yusof, J. F. Zasadzinski, T.-W. Li, 
D. G. Hinks, K. E. Gray, {\em Physica C} {\bf 320},
9 (1999)

\bibitem{miyakawa03} N. Miyakawa, K. Tokiwa, S. Mikusu, J. Zasadzinski, 
L. Ozyuzer, T. Ishihara, T. Kaneko, T. Watanabe, K.E. Gray, {\em Int. J. Mod. Phys. B} {\bf 17},
3612 (2003)

\bibitem{zasadzinski01} J. F. Zasadzinski, L. Ozyuzer, N. Miyakawa, K.E. Gray, 
D. G. Hinks, C. Kendziora, {\em Phys. Rev. Lett} {\bf 87},
067005 (2001); J. F. Zasadzinski, L. Ozyuzer, L. Coffey, K.E. Gray, 
D. G. Hinks, C. Kendziora, {\em Phys. Rev. Lett} {\bf 96},
017004 (2006).

\bibitem{markiewicz} R. S. Markiewicz, C. Kusko, {\em Phys. Rev. Lett} {\bf 84},
5674 (2000)

\bibitem{norman} M.R. Norman, M. Randeria, H. Ding, J.C. Campuzano, {\em Phys. Rev. B} {\bf 52},
615 (1995)


\bibitem{yusof} Z. Yusof, J. F. Zasadzinski, L. Coffey, N. Miyakawa, {\em Phys. Rev. B} {\bf 58},
514 (1998)

\bibitem{won} H. Won and K. Maki, {\em Phys. Rev. B} {\bf 49},
1397 (1993)

\bibitem{zasadzinski03} J. F. Zasadzinski, L. Coffey, P. Romano, Z. Yusof, {\em Phys. Rev. B} {\bf 68},
180504(R) (2003)


\end{references}

\end{document}